# A Numerical Assessment for Predicting Human Development Index (HDI) Trends in the GCC Countries

Mahdi Goldani

m.goldani@hsu.ac.ir

## Abstract:

This study focuses on predicting the Human Development Index (HDI) trends for GCC countries—Saudi Arabia, Qatar, Kuwait, Bahrain, United Arab Emirates, and Oman—using machine learning techniques, specifically the XGBoost algorithm. HDI is a composite measure of life expectancy, education, and income, reflecting overall human development. Data was gathered from official government sources and international databases, including the World Bank and UNDP, covering the period from 1996 to 2022. Using the Edit Distance on Real sequence (EDR) method for feature selection, the model analyzed key indicators to predict HDI values over the next five years (2023-2027). The model demonstrated strong predictive accuracy for in-sample data, but minor overfitting issues were observed with out-of-sample predictions, particularly in the case of the UAE. The forecast results suggest that Kuwait, Bahrain, and the UAE will see stable or slightly increasing HDI values, while Saudi Arabia, Qatar, and Oman are likely to experience minimal fluctuations or slight decreases. This study highlights the importance of economic, health, and educational indicators in determining HDI trends and emphasizes the need for region-specific predictive models to improve accuracy. Policymakers should focus on targeted interventions in healthcare, education, and economic diversification to enhance human development outcomes.

## Introduction

The development demonstrates community welfare. Development can be called the desire to create better conditions and move forward. Unlike the concept of progress, which focuses on the economic dimension, development is a broader concept than economic growth [ 1]. UNDP defines human development as a process of expanding the options for the population in terms of income, health, education, physical environment, and so on. According to BPS, the three basic dimensions as a reference to measure the Human Development Index which includes longevity and a healthy life (a long and healthy life), knowledge, and standard of living (decent standard of living) (fig1).

Fig1. Operationalization of HDI concept

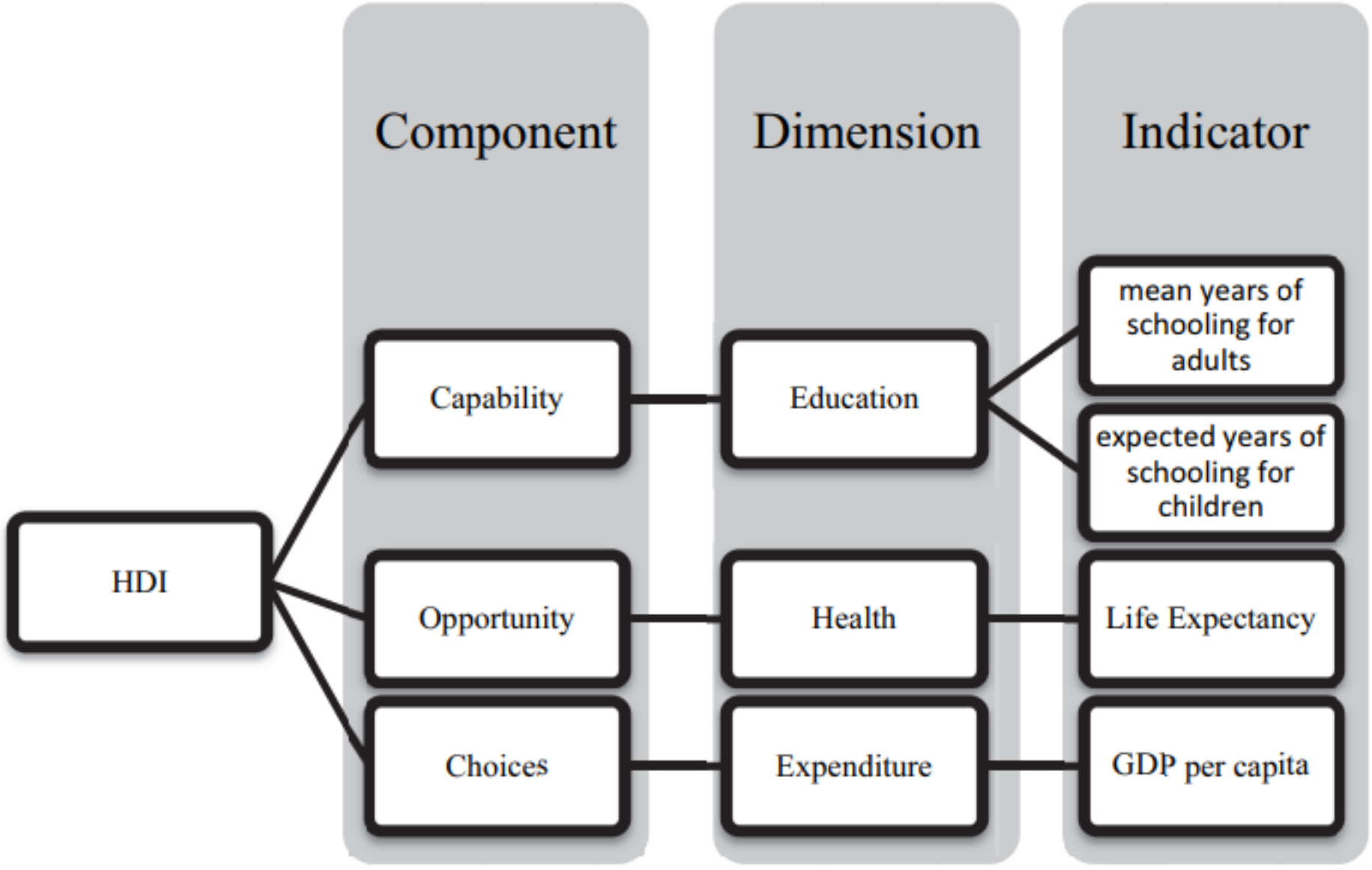

Source: Rama and Yusuf [2]

Human development returns to the main issues societies with different systems and levels of growth[2, 3]. In the Gulf Cooperation Council (GCC) countries, which include Bahrain, Kuwait, Oman, Qatar, Saudi Arabia, and the United Arab Emirates (UAE), HDI values generally reflect high human development levels, although there are variations across these countries. Despite the high HDI scores(fig2), certain structural challenges remain. Economic dependence on oil and gas revenues poses risks to long-term sustainability, and, in some cases, the need to further diversify employment sectors and build skilled labor forces remains a pressing issue.

Fig2. HDI Score (2022)

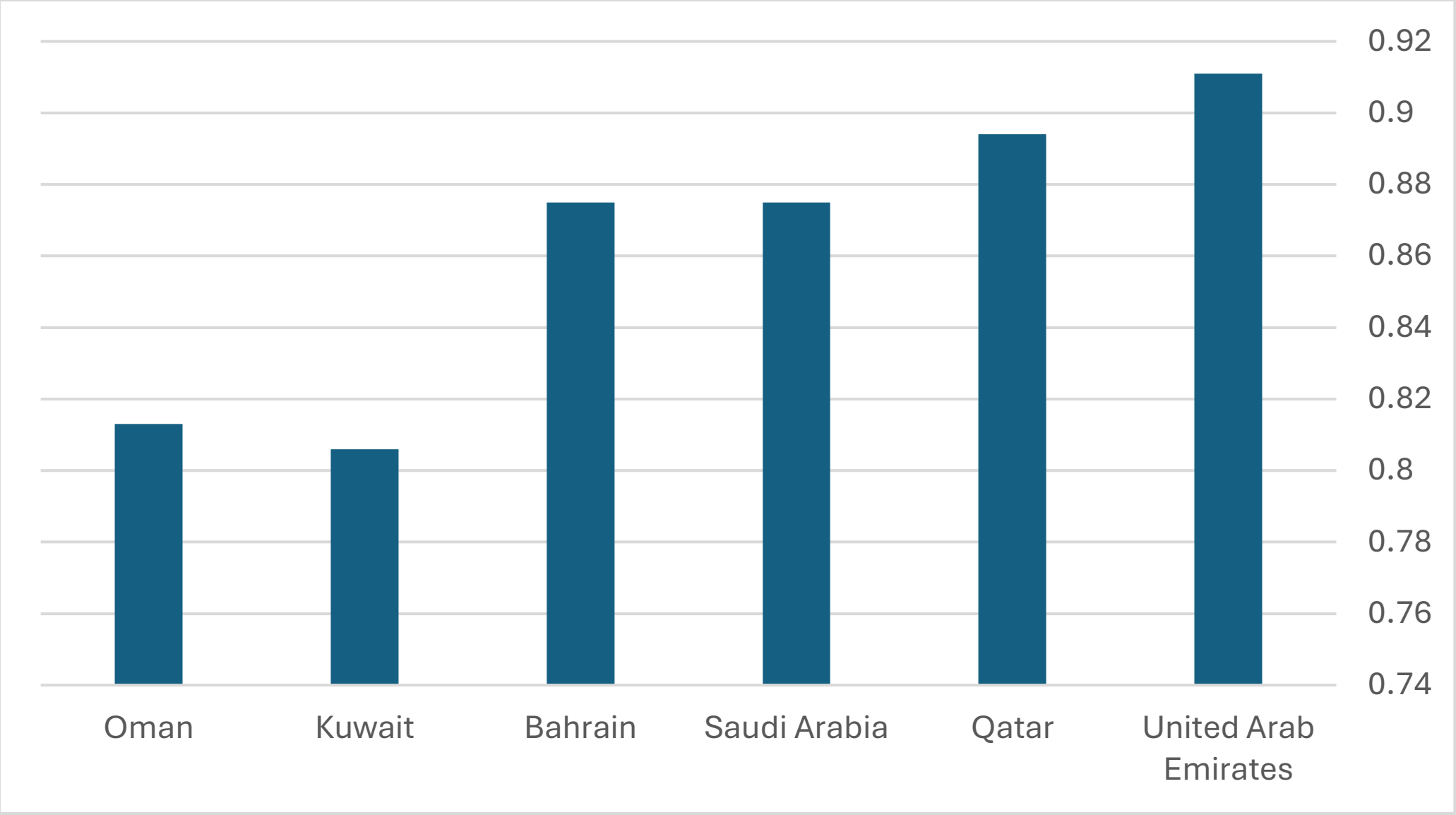

Source: United Nations Development Programmer's (UNDP) (2022)

Therefore, GCC countries need to identify the whole aspect of HDI to address this challenge. These efforts help GCC countries achieve long-term stability by diversifying their economies, building human capital, and adopting sustainable practices. Based on this, the aim of this research is to identify the influencing variables on the human development index in the GCC countries. Then, based on the known variables, a model is made to predict the human development index in the next 5 years for GCC countries.

Various factors directly affect HDI, which are investigated in the literature. These factors can be divided into economic, social, and political categories. Among the economic variables, the GDP growth rate and inflation have the greatest impact on the human development index [4]. Tourovskaia et al. [5] reported that inflation positively and significantly affects HDI, indicating that economic stability and growth have a boosting effect on human development. Meanwhile, some studies showed that inflation and unemployment do not affect the human development index much, and other indicators such as the growth rate of per capita income are among the most effective indicators in determining this index [6]. Yumashev et al [7] demonstrate that the volume of energy consumption not only affects the human development index in a particular country but is also an important factor in determining the level of sustainable development. And also, emissions of carbon dioxide can be effective on HDI as an environmental variable [4]. Bhowmik (2019) argued that there were significant long-run causalities from education expenditure, health expenditure, and GDP per capita to the HDI of SAARC but they had no short-run causalities [8].

Social and demographic variables such as economic variables are factors influencing the human development index. Studies show that population variables and inequality have a significant effect on human development index [3]. Moreover, McCoy et al. [9] imply that improved health and nutritional status may be associated with a higher HDI ranking, thus putting additional focus on the needs of the region with respect to prioritized health interventions. Recently, predictive modeling techniques have evolved as a feasible strategy to estimate HDI based on various predictors. Indeed, as part of his research, it was indicated by Ingber [10] that country level prevalence of low cognitive and socio-emotional scores on ECDI are best modeled using HDI as a predictor. Some research showed that human capital indicators are one of the most relevant indicators to political stability. This issue shows the importance of the role of stability and security law in achieving a high human development index[11]. That kind of modeling is especially useful in the context of the GCC, where early years are crucial for future development of human capital. Furthermore, the study by Hume et al. [12] quantified causes of variation in preterm birth rates among countries with high HDI and thus provided insight into the ways in which specific health outcomes can be modeled to predict HDI.

These findings are indicative of a pathway in which future research might give more consideration to the quantification of additional predictors relevant for GCC countries. Despite the broad research on different health and socioeconomic variables associated with HDI, there are still significant knowledge gaps. Specifically, there is a lack of studies adapted to the unique cultural, economic, and political contexts of the GCC countries. Indeed, predictive modeling studies in the future should focus on localized levels, taking into consideration the peculiar characteristics of these nations. While some literature has utilized different variables of healthcare outcomes and socio-economic factors, comprehensive studies that take in the aspects of environmental factors such as climate change and management of resources into consideration are lacking in the HDI predictions. In this regard, integration of such factors into HDI predictions is quite imperative in showing how the vulnerabilities of the GCC region to environmental changes are crucial to developing resilient human development strategies.

Therefore, the prediction of HDI for the GCC countries becomes quite complex due to the influence of many health, socio-economic, and environmental factors. Though the findings are important and significant associations were traced, region-specific predictive models are still being developed. Such knowledge gaps need to be addressed in order to make effective policy and strategic interventions toward human development enhancement in the GCC.

## Methodology

### Dataset

HDI is a composite index that embodies features relating to human development. These have been devised by the UNDP to measure and compare the level of human development in countries. HDI represents one composite measure, combining three important dimensions: life expectancy at birth-reflecting the health and longevity of the population; mean and expected years of schooling that denote educational attainment and access, respectively; and GNI per capita adjusted for PPP, which provides the living standards. These components are then normalized to ensure comparability and combined into an HDI value by using a geometric mean that ranges from 0 to 1. The closer the value of HDI is to 1, the higher the level of human development.

The data was collected for six GCC countries, namely: Oman, Saudi Arabia, Qatar, Bahrain, the United Arab Emirates, and Kuwait over the period from 1996 to 2022. Additionally, annual data was elicited on life expectancy, educational indicators, and GNI per capita through official government reports, international databases including the World Bank and UNDP, and national statistical offices. The HDI for period and country, accordingly, was calculated using the standard

UNDP methodology of normalization and extreme value adjustments. This ensures that HDI is reliable and consistent over time as a measure of human development and that comparisons of progress in human development can be made across countries. Trends and changes in HDI for such countries have been assessed to review improvements in health, education, and economic conditions over the stipulated period.

The World Bank annually reports 266 indicators for 217 countries and 49 regions. Indicators shed light on the economic, social, political, and environmental situations of every country and area; thus, it can be taken as a complete data set. In this present study, all the indicators of the World Bank have been used to find out which of them resembles political stability and becomes a good explainer of the target value.

## Model

Machine learning (ML) is a powerful tool for making accurate and reliable predictions and, a sub-category of computational intelligence techniques mainly employed for deriving definitive information out of large sets of data for pattern recognition, classification, function approximation, and so forth. With the availability of vast datasets in the era of Big Data, producing reliable and robust forecasts is of great importance [13, 14, 15]. XGBoost is used in this study as an ensemble learning-based algorithm, where a set of base models are combined to create a model that obtains better performance than a single model [16]. It is considered a good method due to its combination of accuracy, efficiency, and flexibility. It surpasses in handling large datasets, managing missing data, preventing overfitting, and offering various tuning options. Its ability to process data quickly, handle imbalanced datasets, and provide interpretable models makes it a go-to choice for many machine learning practitioners. Whether for academic research or industrial applications, XGBoost remains a top choice for building robust and effective predictive models.

## Extreme gradient boosting (XGBoost)

Chen and Guestrin [17] have created Extreme gradient boosting (XGBoost). XGBoost stands for "Extreme Gradient Boosting", where the term "Gradient Boosting" originates from the paper Greedy Function Approximation: A Gradient Boosting Machine, by Friedman [18]. XGBoost is based on the gradient boosting algorithm, a key method in ensemble learning. It combines weak classifiers to create a stronger model, enhancing efficiency and flexibility compared to a single model. By iteratively building decision trees, XGBoost improves classification performance.

Table1. Comparison of XGBoost with Other Prediction Methods in Machine Learning

| Method | Strengths | Weaknesses | Best Use Cases |
|---|---|---|---|
| XGBoost | High accuracy, speed, handles large datasets, built-in regularization to prevent overfitting | More complex to tune, computationally intensive for large hyperparameter grids | Large datasets, competitions, handling missing data, high performance required |
| Random Forest | Good for preventing overfitting, easy to interpret, works well with categorical data | Less efficient on very large datasets, less accurate than boosting algorithms | Smaller or medium-sized datasets, when interpretability is important |
| Support Vector Machines (SVM) | Effective for complex decision boundaries, works well on smaller datasets | Computationally expensive for large datasets, sensitive to feature scaling | Smaller datasets, complex classification problems, high-dimensional spaces |
| Neural Networks | Excels with unstructured data, can model complex relationships | Prone to overfitting, requires large datasets, complex to tune | Unstructured data (e.g., images, text), when deep feature extraction is needed |

Source: [19]

A salient characteristic of objective functions is that they consist of two parts: training loss and regularization term Eq. (1)

$$obj^{(t)} = l(f_t) + \Omega(f_t) \quad (1)$$

In Eq. (1), $f_t$ representing the t-th tree model, $l(f_t)$ is the loss function in the risk prediction, $\Omega(f_t)$is the regular term used to reduce overfitting, which can be expressed as Eq.(2)

$$\Omega(f_t) = \gamma T + \frac{1}{2}\lambda\|\omega\|^2 \quad (2)$$

In Eq. (2), T represents the number of leaf nodes in the t-th decision tree, $\gamma$ and $\lambda$ can decide penalty strength together, $\omega$ representing the weight value on each leaf node. The training loss measures how predictive model is with respect to the training data. A common choice of $L$ is the mean squared error, which is given by Eq.(3)

$$l(f_t) = \sum_i (y_i^t - \hat{y}_i^t)$$
(3)

The prediction results of the model are the weighted sum of all the decision trees, when the t-th iteration is performed, the prediction result can be expressed by Eq(4).

$$\hat{y}_i^{(t)} = \sum_{k=1}^{k} f_{(x_i)=} \hat{y}_i^{(t-1)} + f_t(x_i),\ f_k \epsilon F \quad (4)$$

In Eq. (4), $f_t(x_i)$ represents the t-th tree model, F is the decision tree space, is also the set of all sample risk prediction decision trees. Here $\hat{y}_i^{(t)}$represents the prediction results of the sample I after the t-th iteration, and $\hat{y}_i^{(t-1)}$ represents the prediction results of the previous t-1 trees. Therefore, the objective function can be expressed as the formula for Eq. (5):

$$obj^{(t)} = \sum_{i=1}^{n} l\left(y_i, \hat{y}_i^{(t)}\right) + \Omega(f_t) + costant \quad (5)$$

Over fitting is common and undesirable machine learning methods behavior. Overfitting is the production of an analysis that corresponds too closely or exactly to a particular set of data and may therefore fail to fit additional data or predict future observations reliably. One of the solutions to avoid over fitting is tuning parameters to maximize model performance. There are a bunch of hyperparameter tuning methods. By considering the small dataset, Grid Search is a great option because it exhaustively explores all combinations of hyperparameters.

**Model Evaluation**

Due to the limited sample size, the cross-validation method is chosen to obtain reliable insights. Cross-validation is particularly useful for small datasets because it maximizes the use of available data by repeatedly dividing the dataset into training and validation sets[20].

The mean absolute percentage error (MAPE) is employed to assess the prediction accuracy for both training and testing data. MAPE is calculated as follows:

$$\mathrm{MAPE} = \frac{1}{n}\sum_{i=1}^{n}\left|\frac{A_i - F_i}{A_i}\right| \times 100 \quad (6)$$

where $A_i$ is the actual value and $F_i$ is the forecasted value. This metric provides a straightforward and interpretable measure of forecast accuracy (7).

**Result**

**4-1. feature importance**

The predictors in any machine learning prediction model will certainly act crucially to increase the accuracy of the model. The more properly the predictors can explain the behavior of the target variable, the more accurate the forecast will be. Instead of using research literature, this study has used the large dataset of the World Bank to select the most effective predictor. Since it is less sensitive to sample size and computation wise simpler, according to Goldani [21], feature selection was performed using the Edit Distance on Real sequence (EDR) method. In fig3 the importance of each predictor for each dataset is shown.

Fig3. The importance of features

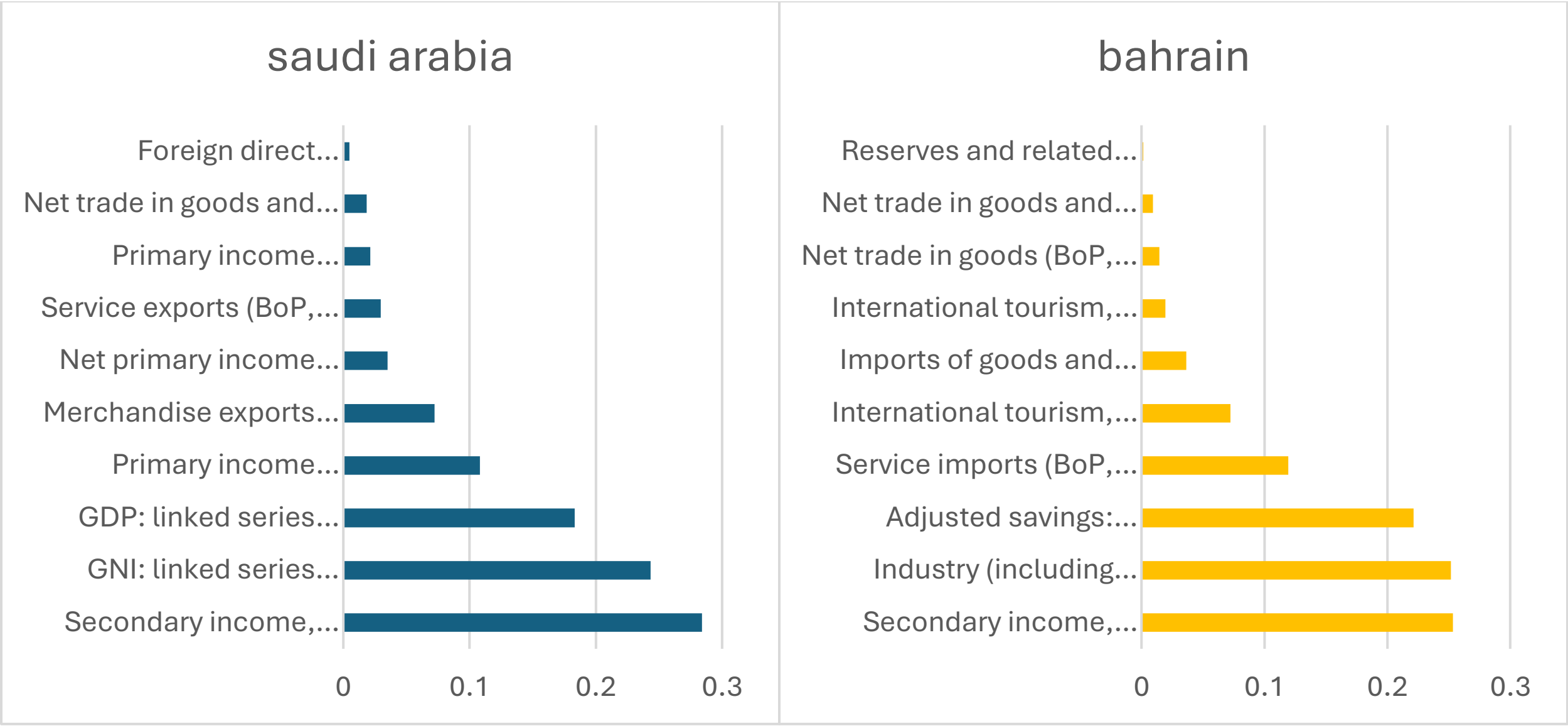

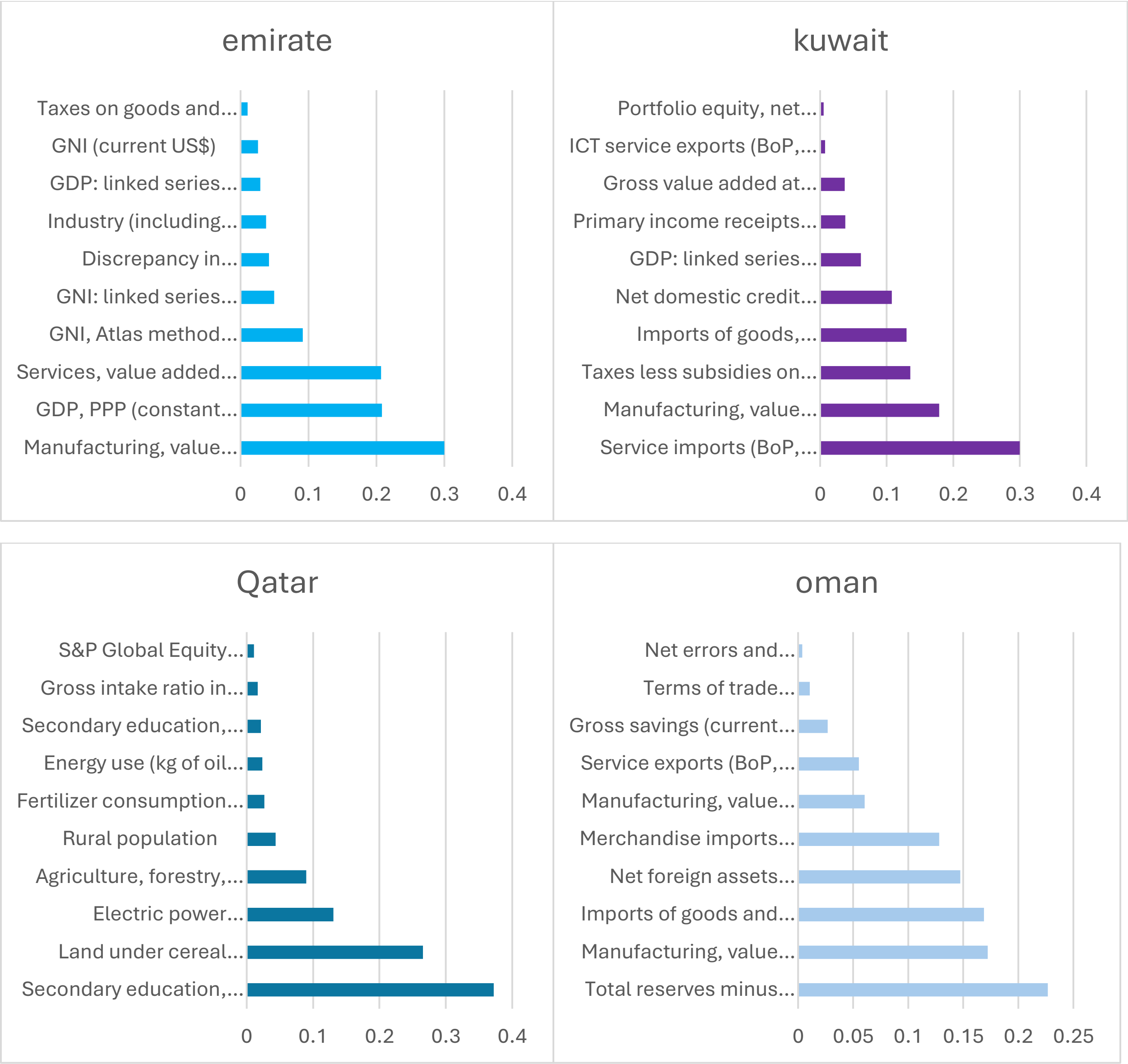

## 4-2. Train/Test Split

The whole predictors and target value were split into training and tests in order to evaluate the performance of the model at each dataset. To improve the performance for the models, building a model within the sample forecast was done. The model within the sample forecast used 1996 to 2018 to estimate the model. Using this model, the forecaster would predict the values from 2018-2022 and compare the forecasted values to the actual known values. The prediction trend of the test data shows that among the six available datasets, the Qatar, Kuwait and UAE datasets better predict the trends of the real data.

Fig4. Actual vs predicted values of test data

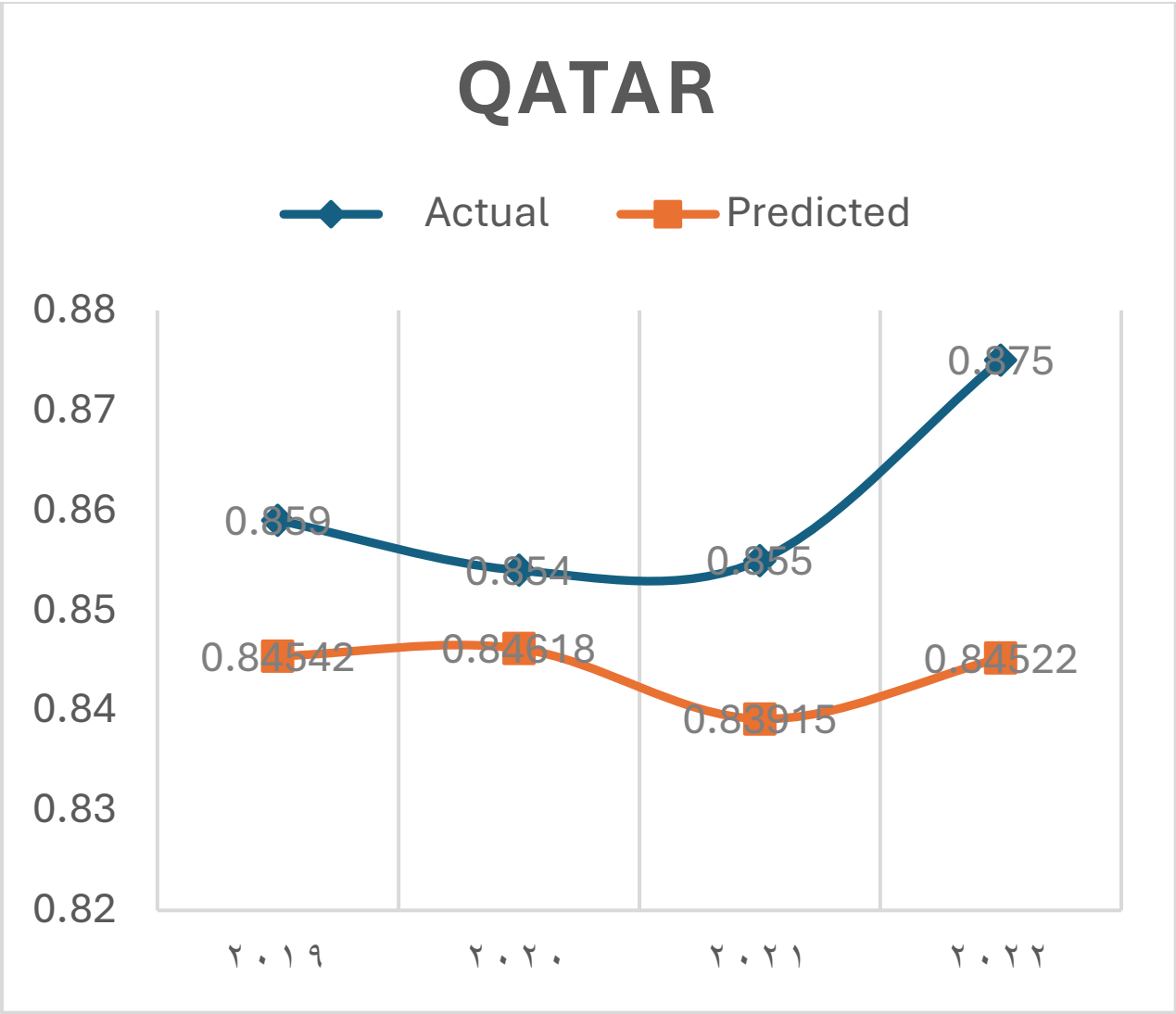

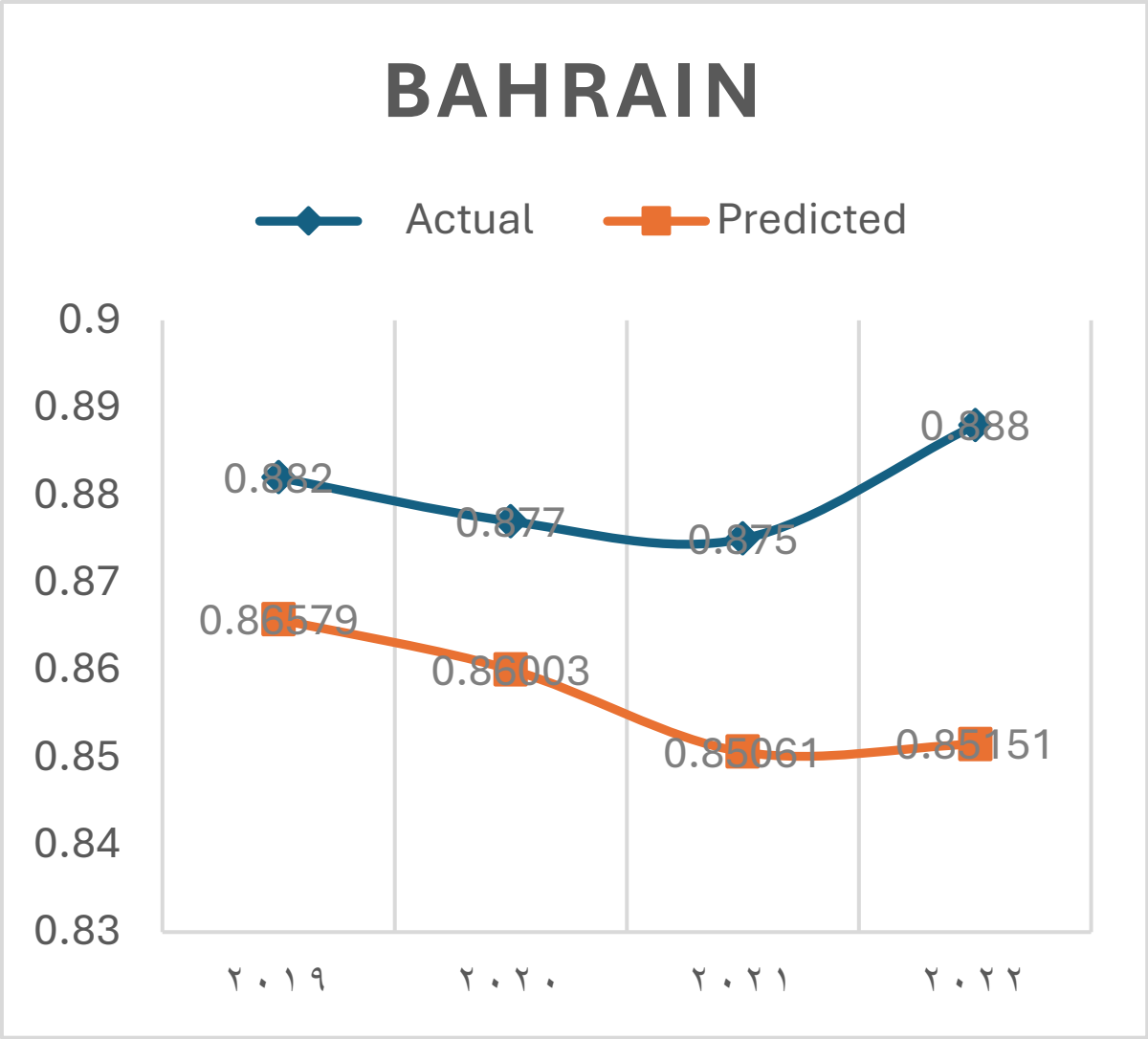

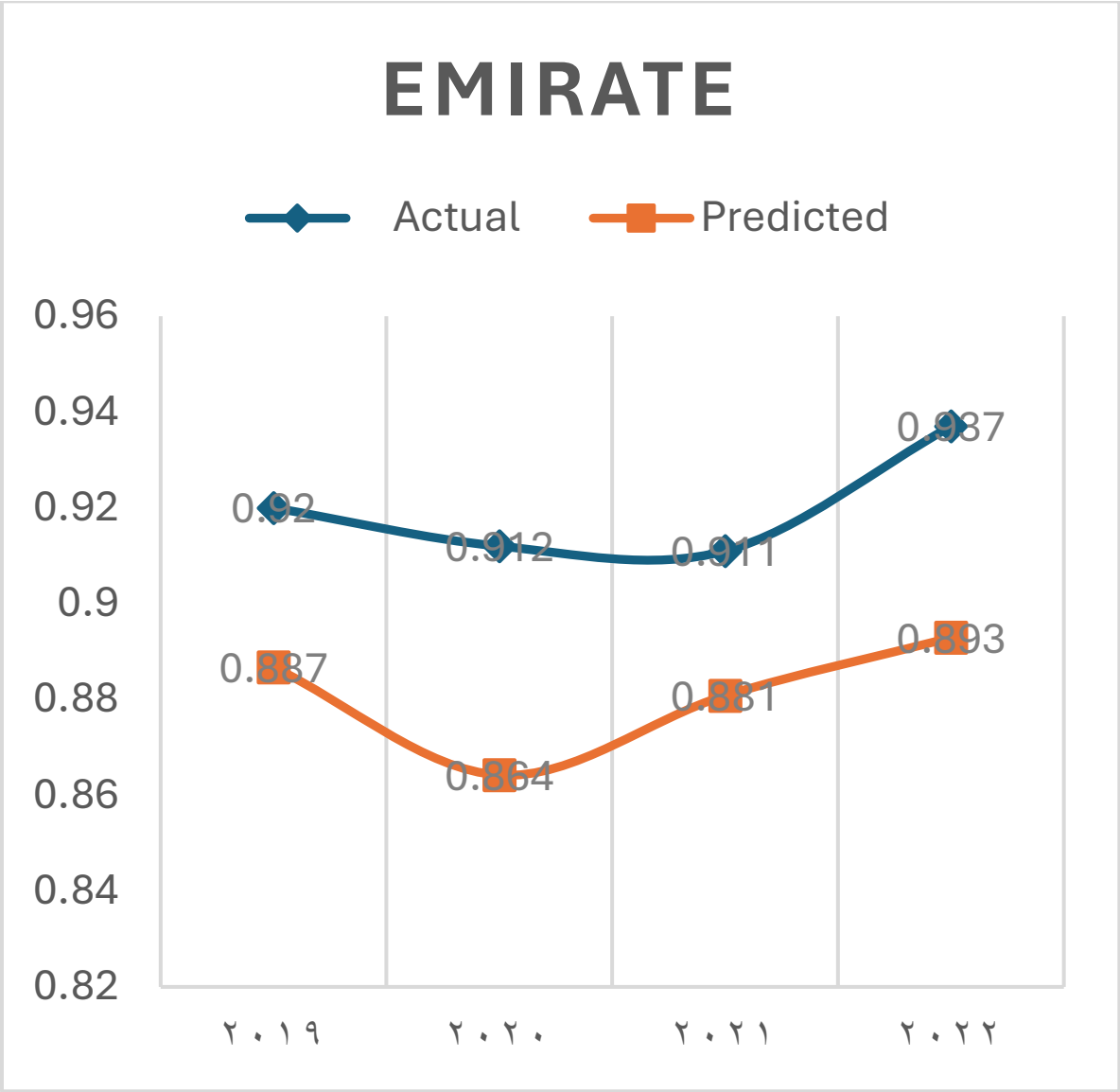

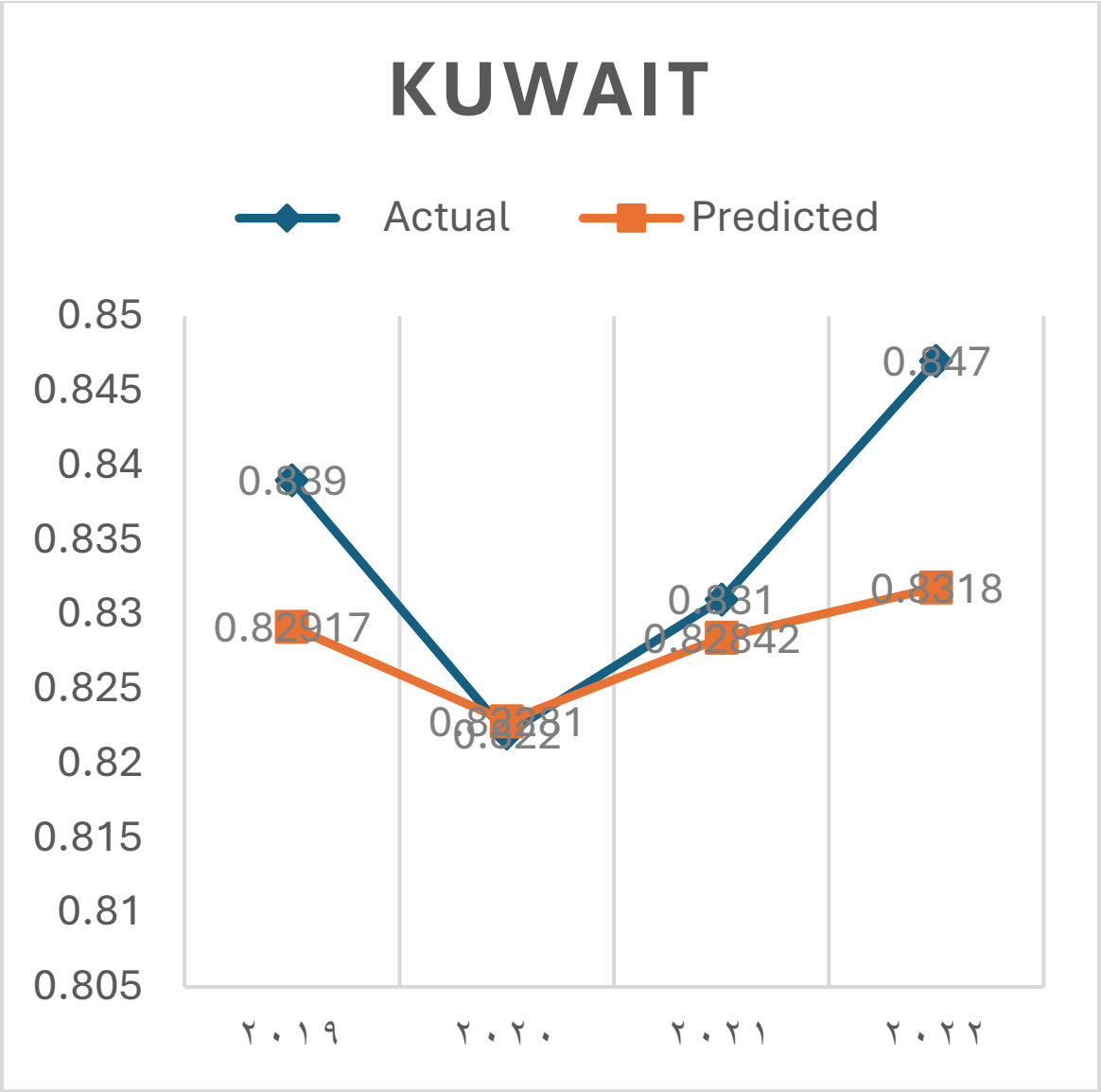

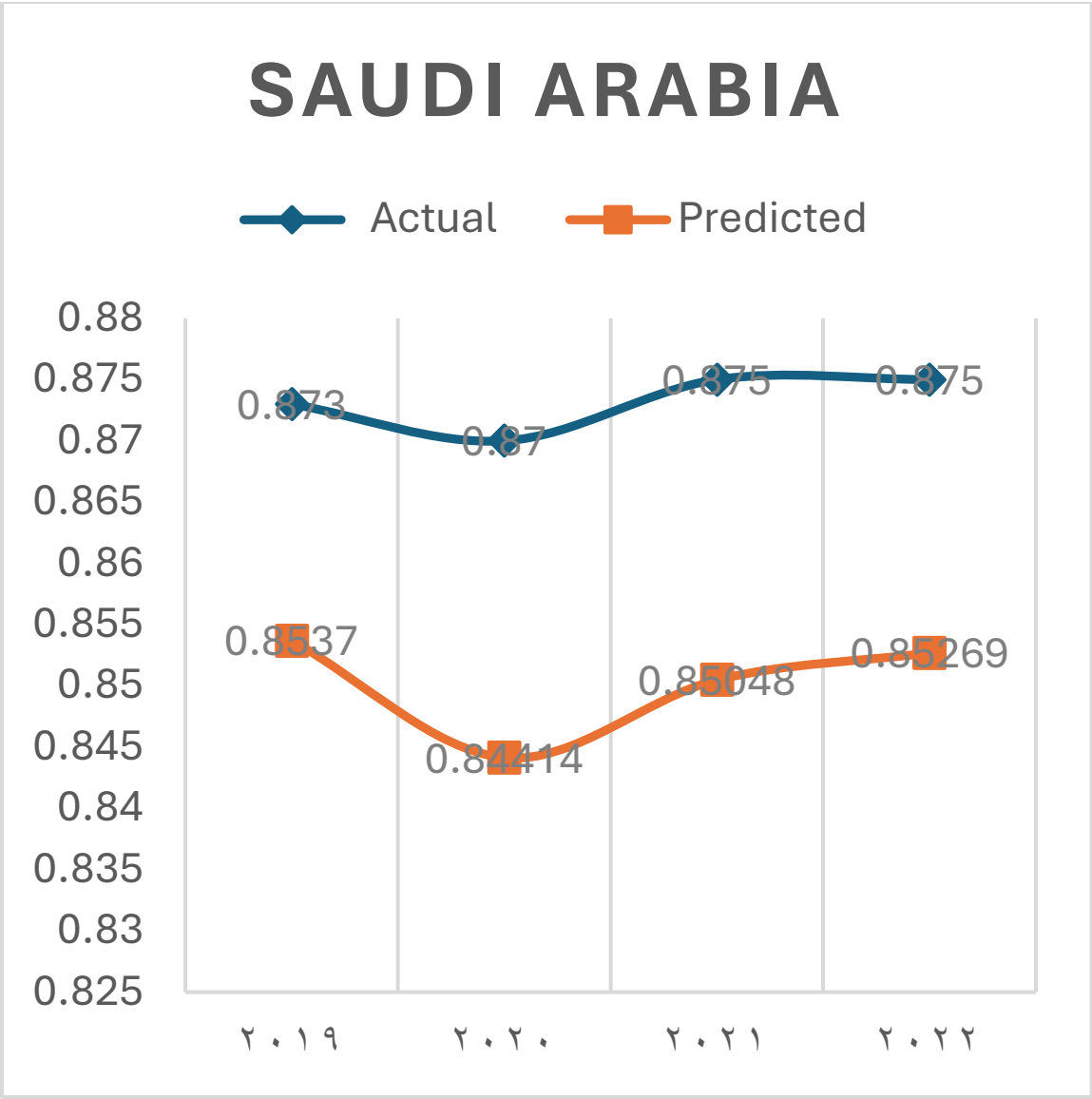

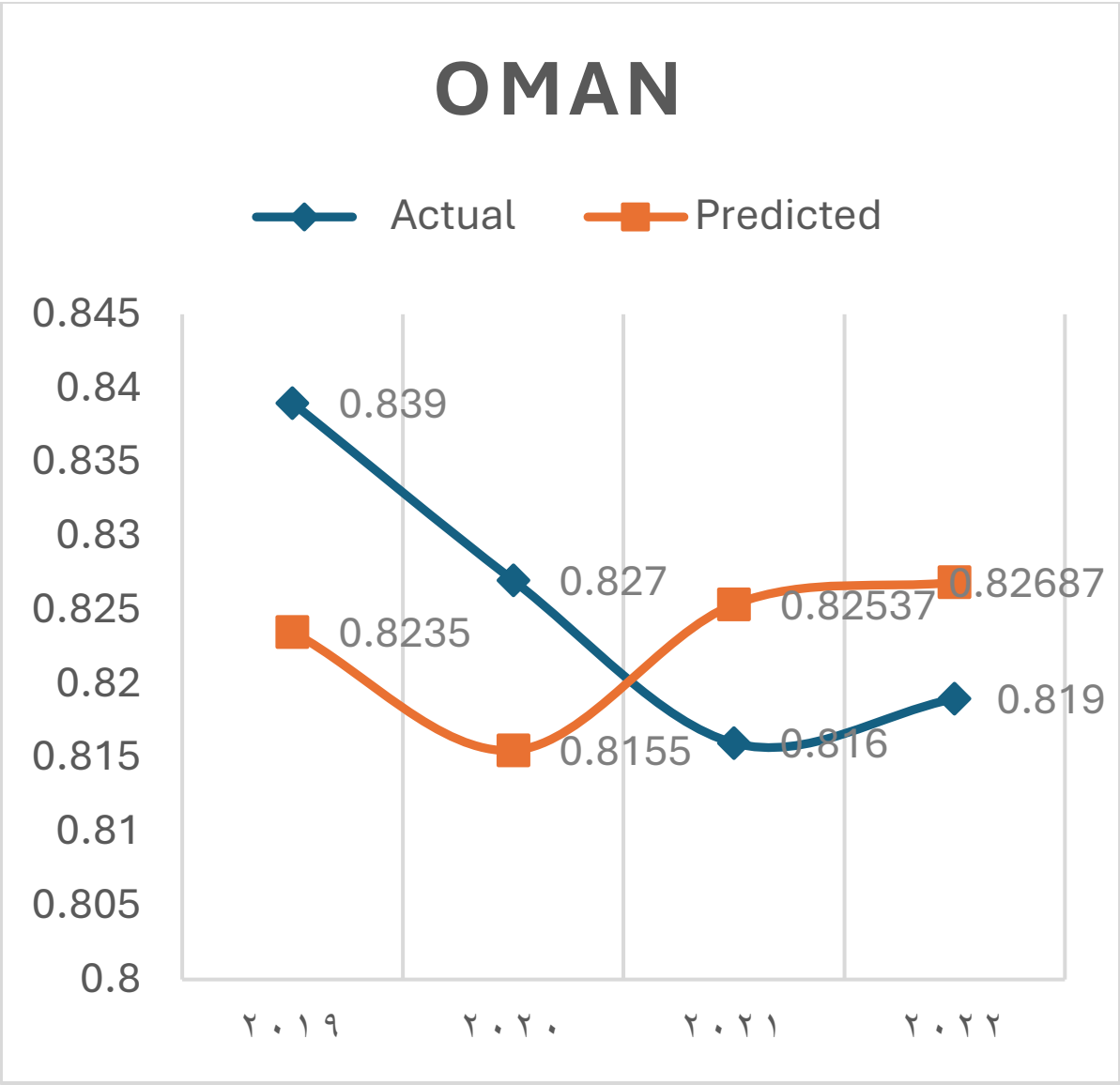

From this, one can comprehend very well that while the model fits quite well for in-sample data, the MAPE shows a sharp increase for out-of-sample data. This is a hint toward potential overfitting issues. Hence, reasons for overfitting must be identified and another approach could be considered for better improvement in generalizing model performance. Kuwait has the best performance among the available datasets. The United Arab Emirates has the worst performance.

Table1. The MAPE of train and test data

| COUNTRIES | | MAPE |
|---|---|---|
| **OMAN** | In-sample MAPE | 0.37% |
| | Out-of-sample MAPE | 1.34% |
| **BAHRAIN** | In-sample MAPE | 0.18% |
| | Out-of-sample MAPE | 2.67% |
| **EMIRATE** | In-sample MAPE | 0.36% |
| | Out-of-sample MAPE | 4.22% |
| **KUWAIT** | In-sample MAPE | 0.19% |
| | Out-of-sample MAPE | 0.84% |
| **QATAR** | In-sample MAPE | 0.32% |
| | Out-of-sample MAPE | 1.94% |
| **SAUDI ARABIA** | In-sample MAPE | 0.44% |
| | Out-of-sample MAPE | 2.63% |

Fig5. The MAPE of dataset

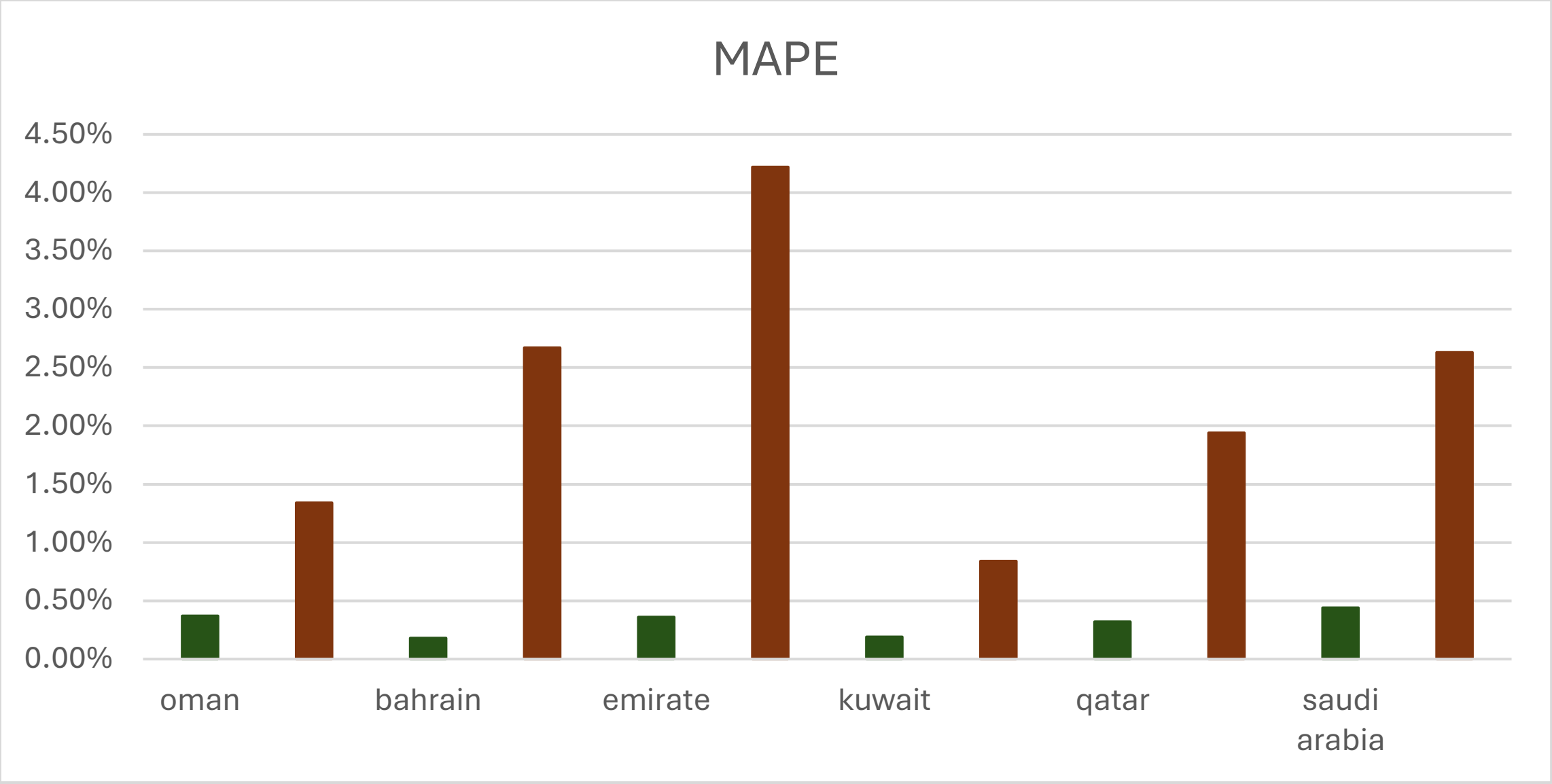

**simulation the predictors**

The next step is to predict the behavior of the political stability index in the six countries under investigation for the next five years. The predictors are simulated for the next five years. The Arima model was used to simulate 5 years of predictors. The historical trends of each time series were used to predict the future trend of them.

**forecast of political stability**

HDI, or the Human Development Index, is a composite measure for assessing key dimensions of human development: life expectancy, education, and per capita income. For most of these countries, HDI assumes stable or rather slightly increased values, meaning further improvement in life expectancy, education, and income levels in the next five years.

Saudi Arabia, Qatar, and Oman experience slight decreases or standstill in their HDI, while Kuwait, Bahrain, and the UAE display stability or a slight increase. The table above indicates that human development in the GCC countries remains on a high level, with just insignificant oscillations from year to year within the forecasted period. More specifically, the leading HDI scores refer to the United Arab Emirates whereas Oman demonstrates the lowest indicators within this group.

Fig6. forcasting of human development index

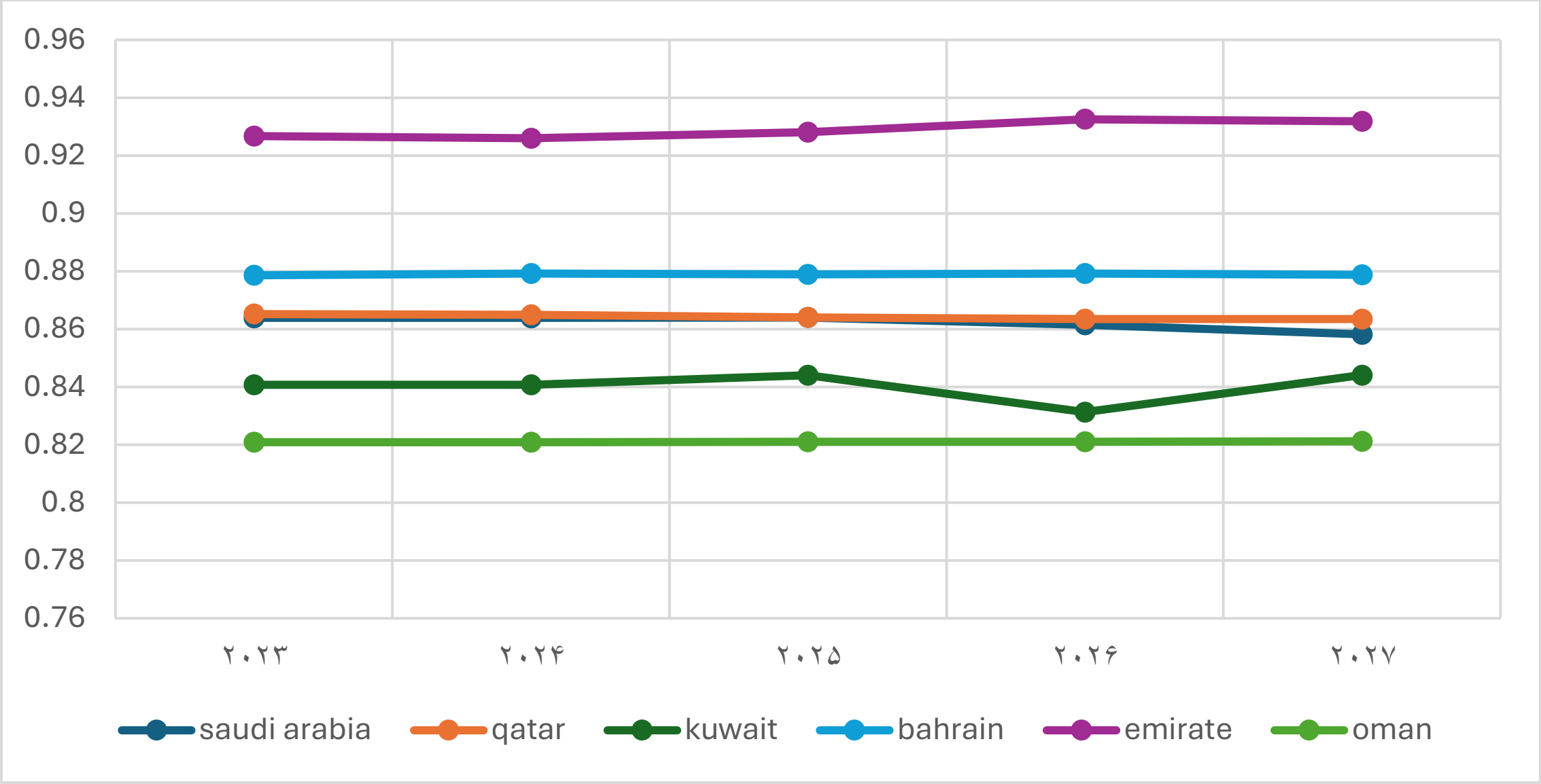

**Conclusion**

The predictive modeling of the Human Development Index (HDI) for the six GCC countries Saudi Arabia, Qatar, Kuwait, Bahrain, the United Arab Emirates, and Oman provides valuable insights into future human development trends. The analysis indicates that these countries, despite their varying economic structures and socio-political contexts, are likely to experience relatively stable HDI scores over the forecasted five-year period (2023-2027).

Countries like Kuwait, Bahrain, and the UAE show stability or slight improvements in their HDI scores, signifying consistent progress in health, education, and income levels. On the other hand, Saudi Arabia, Qatar, and Oman exhibit slight decreases or standstill in their HDI predictions, suggesting that while their human development remains high, growth may slow in the coming years. The United Arab Emirates consistently leads with the highest HDI among the countries studied, highlighting its strong performance across all HDI dimensions, including education, life expectancy, and income. Oman consistently lags in HDI performance compared to its GCC counterparts, suggesting that it may face more significant challenges in improving its socio-economic and health indicators.

The machine learning model used in this study, specifically XGBoost, shows strong predictive performance for in-sample data. However, the model exhibited some overfitting when tested with out-of-sample data, particularly in the case of the UAE. While Kuwait had the best predictive accuracy, the UAE demonstrated the most significant gap between predicted and actual HDI values, indicating potential improvements needed in the model to better capture the complexity of human development in these nations.

The findings underscore the importance of including diverse health and socio-economic variables in predictive models. For instance, indicators like preterm birth rates and iron deficiency anemia showed strong correlations with HDI in previous studies, and their integration into future models could enhance prediction accuracy. The unique cultural, political, and economic environments of the GCC countries suggest that region-specific models could provide more accurate forecasts. Future research should focus on the integration of environmental factors, such as climate change and resource management, which are crucial for the resilience of human development strategies in this region.Policymakers in the GCC countries should consider targeted interventions in healthcare, education, and economic diversification to improve their HDI scores. Countries like Saudi Arabia and Oman, in particular, could benefit from policies that address their relatively lower HDI performance.

Overall, while the GCC countries continue to rank highly in terms of human development, their future progress will depend on strategic policies aimed at addressing health, education, and economic challenges specific to the region.

**Refrences**